\documentclass[a4paper, amsfonts, amssymb, amsmath, reprint, showkeys, nofootinbib, twoside, superscriptaddress]{revtex4-1}
\usepackage[english]{babel}
\usepackage[utf8]{inputenc}
\usepackage{verbatim}
\usepackage[colorinlistoftodos, color=green!40, prependcaption]{todonotes}
\usepackage{amsthm}
\usepackage{mathtools}
\usepackage{physics}
\usepackage{xcolor}
\usepackage{graphicx}
\usepackage[left=23mm,right=13mm,top=35mm,columnsep=15pt]{geometry} 
\usepackage{adjustbox}
\usepackage{placeins}
\usepackage[T1]{fontenc}
\usepackage{lipsum}
\usepackage{csquotes}
\usepackage[pdftex, pdftitle={Article}, pdfauthor={Author}]{hyperref} 

\begin{document}
\title{Spatially Resolved Thermoelectric Effects in Operando Semiconductor-Metal Nanowire Heterostructures}

\author{Nadine G\"achter}
    \email[Correspondence email address: ]{nadine.gaechter@unibas.ch}\affiliation{IBM Research - Zurich, 8803 R\"uschlikon, Switzerland}

\author{Fabian K\"onemann}
    \affiliation{IBM Research - Zurich, 8803 R\"uschlikon, Switzerland}

\author{Masiar Sistani}
    \affiliation{Institute of Solid State Electronics - TU Wien, 1040 Vienna, Austria}

\author{Maximilian G. Bartmann}
    \affiliation{Institute of Solid State Electronics - TU Wien, 1040 Vienna, Austria}

\author{Marilyne Sousa}
    \affiliation{IBM Research - Zurich, 8803 R\"uschlikon, Switzerland}

\author{Philipp Staudinger}
    \affiliation{IBM Research - Zurich, 8803 R\"uschlikon, Switzerland}

\author{Alois Lugstein}
    \affiliation{Institute of Solid State Electronics - TU Wien, 1040 Vienna, Austria}

\author{Bernd Gotsmann}
    \email[Correspondence email address: ]{bgo@zurich.ibm.com}
    \affiliation{IBM Research - Zurich, 8803 R\"uschlikon, Switzerland}

\date{\today} 

\begin{abstract}
The thermoelectric properties of a nanoscale germanium segment connected by aluminium nanowires are studied using scanning thermal microscopy. The germanium segment of 168\,nm length features atomically sharp interfaces to the aluminium wires and is surrounded by an Al$_2$O$_3$ shell. The temperature distribution along the self-heated nanowire is measured as a function of the applied electrical current, for both Joule and Peltier effects. An analysis is developed that is able to extract the thermal and thermoelectric properties including thermal conductivity, the thermal boundary resistance to the substrate and the Peltier coefficient from a single measurement. Our investigations demonstrate the potential of quantitative measurements of temperature around self-heated devices and structures down to the scattering length of heat carriers.
\end{abstract}


\maketitle

Thermal transport and energy conversion at the scale of micrometers to nanometers is a fascinating topic of research. The carriers of heat and charge as well as their transport mechanisms have characteristic length scales in this regime. Consequently, a multitude of effects can be studied and ultimately exploited\cite{Pop2010,Shi2012_nanostructures}. Thermoelectric energy conversion, for example, has been predicted and shown to occur at enhanced efficiency in micro- and nanoscale structures\cite{Hicks1993,Shi2012_nanostructures, Pop2010}. Particularly interesting realizations are axial and radial nanowire heterostructures, combining effects of reduced dimension both in the transport direction and perpendicular to it\cite{Sistani2018,Swinkels_2018}. The transition from an extended thermoelectric material to a finite length along the transport direction, however, has not yet been fully explored. For example, the influence of interfaces between a thermoelectric material and metallic contacts or the transition between the conventional Peltier effect and thermionic emission may still hold some insights in the experimental realization\cite{Shakouri1998}.

One of the challenges in researching this area is the difficulty in performing nanoscopic heat transport measurements\cite{Pop2010}. Thermal contact resistances dominating at small length scales translate into systematic challenges for measuring the local temperature or heat flux in a structure. The thermal contacts to nanowire samples, for example, have turned out to be a major issue\cite{Swinkels_2018,Menges_nature}. It has been proposed that this can be mitigated using extended measurement series for example of thermal transport as a function of length, sometimes called transmission line method\cite{Vakulov2020} or related\cite{Kim2015}.  Apart from being time-intensive, these solutions rely on other assumptions such as the ability to fabricate reproducible test structures with the same contact resistances to electrodes or thermometers. For semiconducting nanowires this is, oftentimes, a major difficulty\cite{Wagner2018}.

Motivated by the recent success of using scanning thermal microscopy (SThM)\cite{gomes2015}, we here describe the development of a method to systematically extract thermal and thermoelectric transport properties from the spatial temperature information. In contrast to transport measurements using external heaters and thermometers, here we analyse the temperature distribution in self-heated nanostructures to extract information through fitting with appropriate models. While thermal conductivity measurements with SThM are often hampered by the thermal resistance between probing tip and sample, we apply periodic self-heating to circumvent this problem and are able to measure the sample temperatures\cite{Menges_nature}. In this way, we can not only mitigate but also quantify the effect of thermal contact resistances inside and around the device. Thereby, we can address length scales not readily accessible using other methods (such as a segment and its interfaces within a nanowire). Additionally, the temperature distribution is measured in its topographic context by atomic force microscopy (AFM). It is by that possible to align and compare scans under different operating currents. Also, the method is independent on the geometry and material of the device. This means, that there is no need for altering the sample architecture for the thermal measurements. 

\begin{figure}[tb] 
	\centering 
	\includegraphics[width=1.0\columnwidth]{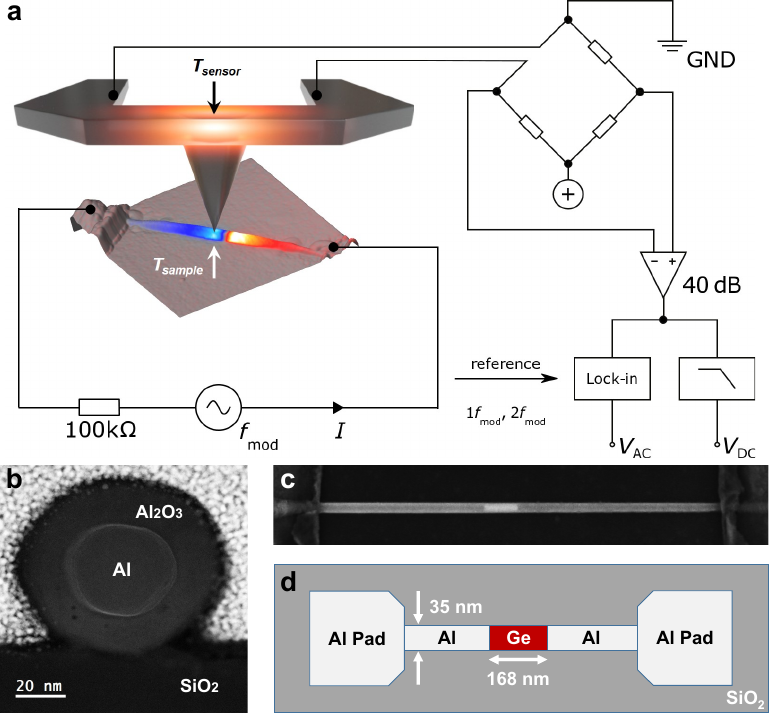} 
    \caption{a) Schematic of the SThM set-up. A slow AC-bias is applied over the device to induce an oscillating sample temperature $T_{\rm sample}$ which is thermally coupled via the tip to the heated sensor. The signal is de-modulated with a lock-in amplifier at the modulation frequency $f_{\rm mod}$ and $2f_{\rm mod}$ for measuring the temperature increase due to the Peltier and Joule effect respectively. The simultaneous measurement of the AC- and DC-signal allows to eliminate the thermal resistance between sample and tip and to infer $T_{\rm sample}$. b)-d) The single crystalline Al-Ge nanowire is electrically connected over Al electrodes and lies on SiO$_2$ substrate. It is naturally integrated in a back-gated FET set-up. b) shows a dark field scanning transmission electron microscope (DF-STEM) image of the nanowire cross-section, c) a SEM image and d) a sketch of the device.}
    \label{fig:device} 
\end{figure}

There are examples of interpreting the temperature field of nanowire structures using SThM to extract thermal conductivity\cite{Menges_nature, konemann2019, konemann2020, Shi2009_sthm}. Here, we extend the method to extract the thermal conductivity of a thermoelectric material and the metal leads, the thermal interfaces to a substrate, and the Peltier coefficient from the observation of the temperature fields. The nanowire device consists of a germanium (Ge) segment which is monolithically integrated in a single crystalline aluminium (c-Al) nanowire with atomically sharp interfaces\cite{sistani2017} and a diamter of 35\,nm. A dark field scanning transmission electron microscope (DF-STEM) image of the nanowire cross-section, a SEM image and a sketch of the device are shown in Figure \ref{fig:device}(b-d). Further information on the fabrication, electrical and structural characterization of the device can be found in the supplementary information and in \cite{Elhajouri2019, Brunbauer_2016, Kral2015}. The thermoelectric figure of merit can thus be calculated from a single measurement collecting the complementary information from the Joule and the Peltier signals. The analysis concentrates on the temperature profile in the lateral direction of the wire for the different operating currents.

\section{Experimental}

\subsection{Thermal imaging with the SthM}
The present SThM setup is operated in a highly shielded lab environment\cite{NFL} in a high vacuum chamber at a pressure of $10^{-6}$\,mbar. The measured thermal maps have a spatial resolution below 10\,nm and mK sensitivity\cite{Menges_nature}. The technique is based on an AFM, where the tip is operated in contact with the sample. The silicon-based temperature sensor is located at the base of the tip and heated to a temperature $T_{\rm sensor} = 273\,^{\circ}$C, around which the electrical resistance of the sensor depends linearly on the temperature. When the heated tip is put in contact with the sample, the temperature difference induces a heat flow between the sensor and the sample. A change in sample temperature leads to a change in the heat flux that is detected through a small change in sensor temperature. At the same time, the sample temperature is not expected to be influenced beyond the uncertainty of the method as the thermal resistances within the sample are orders of magnitude lower than the thermal resistance between tip and surface. The nanoscopic contact between tip and sample has typically a resistance of more than $ 10^7-10^8$\,W/K compared to the cantilever leads with a thermal resistance of $R_{\rm cl} \approx 2\cdot10^{5}$\,W/K. This difference prevents the sensor from equilibrating with the sample temperature\cite{Menges2016_instruments}.

A major difficulty in the determination of the sample temperature $T_{\rm sample}$ lies in quantifying the thermal resistance of the tip, the tip-sample contact and the spreading into the substrate. They are all summarized in $R_{\rm ts}$, which then depends on the surface material, the precise shape of the tip and largely on the size of the touching point that changes with the surface granularity and the topography of the sample. To overcome this problem, the nanowire is operated with an electrical AC-bias of frequency $f_{\rm mod}=1234$\,Hz that creates a continuously modulated temperature field. $R_{\rm ts}$ is then eliminated in the equations by simultaneous measurement of the time-averaged sensor signal $\Delta V_{\rm DC}$ and the de-modulated sensor signal amplitude $\Delta V_{\rm AC}$. An illustration of the set-up is depicted in Figure \ref{fig:device}a. The sample temperature difference to ambient temperature, $\Delta T_{\rm sample}$, for linear devices is then given by\cite{Menges_nature}
\begin{equation}
\Delta T_{\rm sample} = \Delta T_{\rm sensor} \cdot \frac{\Delta V_{\rm AC}}{\Delta V_{\rm DC} - \Delta V_{\rm AC}}.
\label{eq:sampletemp}
\end{equation}
An additional feature of the lock-in detection is the possibility to separate the temperature change that is caused by Joule heating and the Peltier effect, respectively. Joule heating, on the one hand, is proportional to the dissipated electrical power and the signal appears at $2f_{\rm mod}$. On the other hand, the Peltier effect is directly proportional to the current and produces a thermal signal at $1f_{\rm mod}$.

\section{Results and analysis}
\begin{figure*}[t] 
	\centering
	\includegraphics[width=2.0\columnwidth]{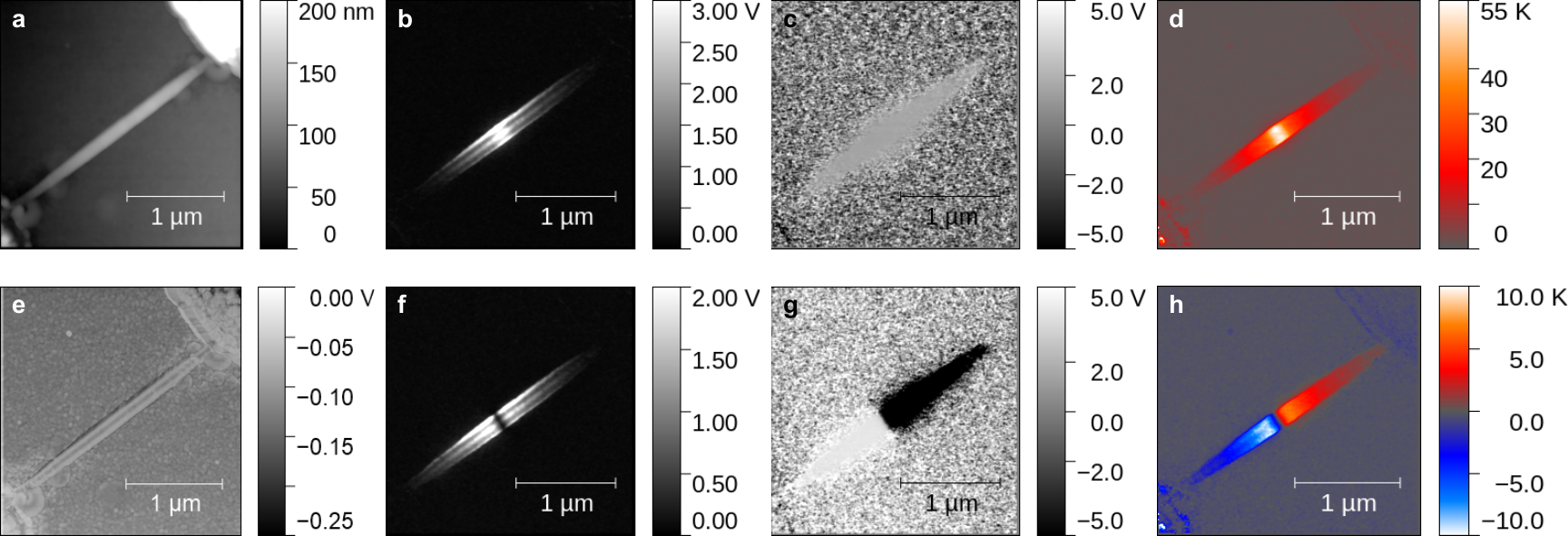} 
	\caption{Scanning thermal microscopy data of the wire operating with an AC-current of amplitude $I = 26\,\mu$A.
	\newline a) AFM topography. b) Thermal signal amplitude taken at 2\,$f_{\rm mod}$ in response to Joule heating. c) Phase of the 2\,$f_{\rm mod}$ signal. d) Joule thermal map calculated using equation\,\ref{eq:sampletemp} and the scans in b) and e). e) DC-thermal measurement proportional to the thermal resistance of the tip-sample contact. f) Thermal signal amplitude taken at 1\,$f_{\rm mod}$ in response to Peltier heating/cooling. g) Phase of the 1\,$f_{\rm mod}$ AC signal, where $\pm$5\,V equals $\pm$90$^\circ$. h) Peltier thermal map calculated using equation\,\ref{eq:sampletemp} and the scans in e), f) and g).}
	\label{fig:sthmmaps} 
\end{figure*}

\subsection{Temperature maps of self-heated nanowires}
Figure \ref{fig:sthmmaps} shows the results from a SthM scan at an operating current of 26\,$\mu$A. (a) is the sample topography from the SThM measurement of the operating nanowire with a spatial resolution of 10\,nm. The inferred temperature fields that are induced by Joule heating $\Delta T_{\rm Joule}$ and the Peltier effect $\Delta T_{\rm Peltier}$ are represented in (d) and (h). Using equation \ref{eq:sampletemp}, they are calculated  from the DC thermal signal (e), and the AC signals at $2f_{\rm mod}$ (b) and $f_{\rm mod}$ (f), respectively. The phase signals (c, g) are recorded with the lock-in detector and show the expected behavior for an approximately linear resistor.

\begin{figure*}[t]
	\centering
	\includegraphics[width=2.0\columnwidth]{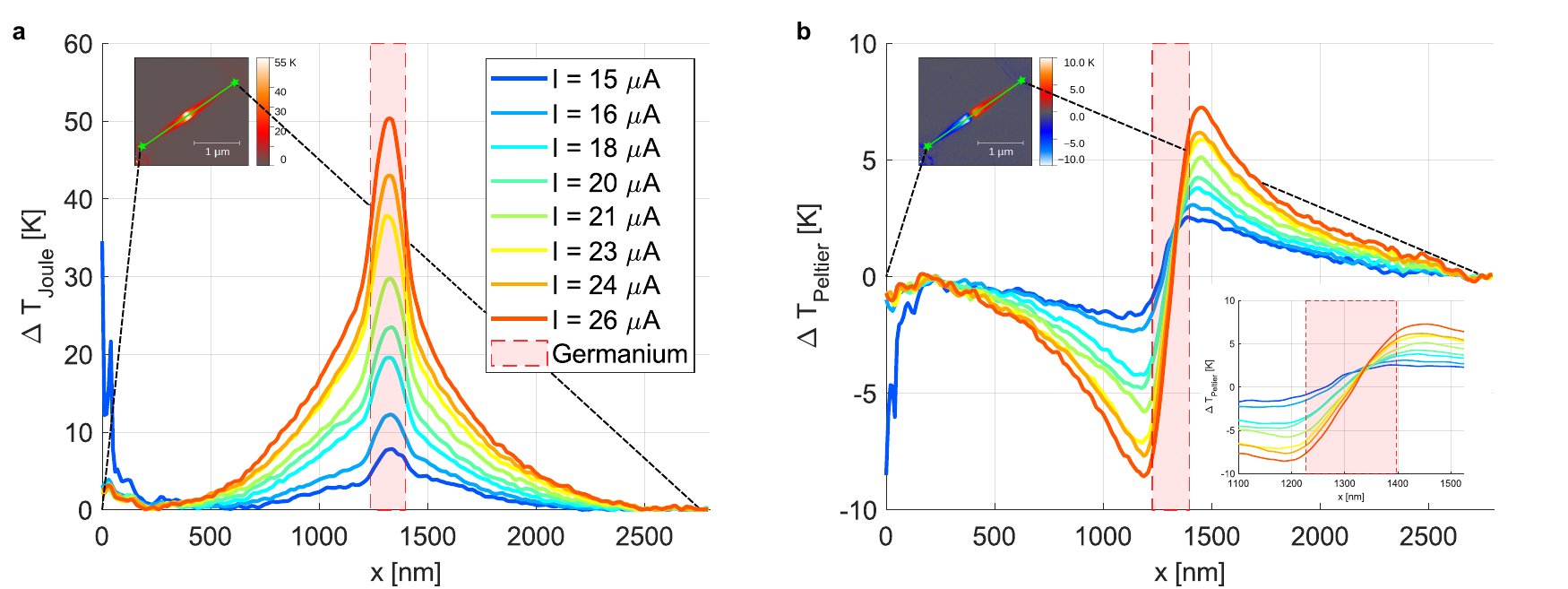}
    \caption{Temperature profiles along the wire with the Ge-segment shaded in red. a) The temperature change induced by Joule heating $\Delta T_{\rm Joule}$ as a function of the position $x$ along the wire, for the operation under increasing voltage biases. b) The temperature change due to the Peltier effect $\Delta T_{\rm Peltier}$ as a function of the position $x$ along the wire operating at different currents $I$ with a zoom on the area of the energy barrier.}
	\label{fig:SThMProfiles} 
\end{figure*}

The following analysis concentrates on the line-profiles extracted from the two-dimensional temperature maps of Figure \ref{fig:sthmmaps} (d) and (h) and according ones for other currents between 13 and 26\,$\mu$A. The temperature profiles along the long axis of the nanowire are shown in Figure \ref{fig:SThMProfiles} for both Joule and Peltier signals.

\subsection{Analysis based on a 1D heat diffusion model}
We analyze the data through fitting the experimental temperature profiles to a model based on the one-dimensional heat diffusion equation. It consists of the following contributions: (1) a heat diffusion term, (2) a Joule heating term, proportional to the square of the current $I$, (3) a Peltier heating term, linearly proportional to the current, (4) a term for heat-loss to the substrate, proportional to the temperature difference between sample and substrate and (5) a term related to the heat capacity $C_V$: 
\begin{equation}
\kappa A\frac{\partial^2T}{\partial x^2} + I^2 \cdot \frac{\rho}{A} + I\cdot \pi - g(T-T_{\rm ambient}) = C_V A\frac{\partial T}{\partial t},
\label{eq:diffusion}
\end{equation}
where $T$ and $T_{\rm ambient}$ are device and ambient temperature, $A$ the nanowire's cross sectional area, $\kappa$ and $\rho$ the material dependent thermal and electric conductivities and $g$ the substrate coupling constant counting the heat loss of the wire per unit length. In this context $\pi$ is the local Peltier coefficient per unit length, as discussed below. The nanowire is slightly tapered on both sides near the metal electrodes. Nevertheless, in the regions of significant temperature rise, the diameter is constant and a position-independent $\kappa$, $A$ and $g$ is justified. A similar diffusion equation model has been used to analyse SThM data of nanowires previously\cite{Shi2009_sthm, Menges_nature, konemann2020}. However, there are underlying assumptions that need to be justified for each system under study:

Firstly, the radial temperature distribution can be neglected. This assumption works well for wires with high aspect ratio as in this case. For this composite system we further require that the thermal interface resistance between the core and Al$_2$O$_3$-shell is smaller than that between wire and substrate. While this is a plausible assumption (see below), our study is limited in studying lateral intra-wire effects.

Secondly, the heat spreading into the substrate from a line heat source has no analytical solution. To be able to simplify this into a single thermal conductance term per unit length between a nanowire and the substrate ($g \Delta T$) requires, that the thermal interface resistance to the substrate is larger than the spreading resistances within the wire and the substrate. As a first indication, we observe a discontinuity in the horizontal temperature profile between wire and substrate. Figure \ref{fig:fits}a shows that the temperature rise along a line section perpendicular to the wire is approximately constant across the nanowire surface and also five to ten times larger than on the substrate. Note, that the wire appears to have a width of 150\,nm (instead of the actual 65 \,nm) due to the well-known convolution effects with the tip shape. In the convolution area and due to the finite diameter of the tip-surface the temperature distribution appears smeared out over some tens of nanometers in this line scan, in contrast to a higher resolution along the nanowire axis.

Thirdly, equation \ref{eq:diffusion} implies that the transport is essentially diffusive. This is a well established assumption for thermal transport in nanowires of these dimensions with one notable exception where reduced dimensions lead to a ballistic regime\cite{Vakulov2020}. However, in this study the observed temperature profiles within the segments are typical for diffusive transport and cannot be explained using ballistic effects. They can be observed only around the interfaces and boundaries, where the spatial resolution of the scan is higher than the carrier scattering length. Near the Ge-Al interfaces, the thermalization lengths of non-equilibrium charge carriers lead to a spatial distribution of the thermoelectric effect\cite{konemann2020, Lake1992, Shakouri1998}. The effective Peltier coefficient is then the integration over the spatial distribution of the local Peltier coefficient
\begin{equation}
\Pi = \int \pi(x) dx.
\label{eq:petcoefficient}
\end{equation} 

Finally, the AC-modulation period of the driving current is orders of magnitudes shorter than any thermal time constant of the system. Therefore, the scan is taken at steady state and the last term of equation\,\ref{eq:diffusion} is negligible.

For a system whose electric response is sufficiently linear, the total temperature distribution is disentangled to $T(x,t) = T_{\rm ambient} + \Delta T_{\rm Peltier}(x)\sin{(2\pi f t)} + \Delta T_{\rm Joule}(x)\sin{(4\pi f_{\rm mod} t)}$ for the current $I(t) = I_0\sin{(2\pi f_{\rm mod} t)}$. The respective thermal profiles are, then, described independently by  
\begin{equation}
\kappa A\frac{\partial^2}{\partial x^2}\Delta T_{\rm Joule} + I_0^2 \cdot \frac{\rho}{A} -  g\cdot \Delta T_{\rm Joule} = 0
\label{eq:diffusionJ}
\end{equation}
and
\begin{equation}
\kappa A\frac{\partial^2 }{\partial x^2}\Delta T_{\rm Peltier} + I_0\cdot \pi - g\cdot \Delta T_{\rm Peltier} = 0.
\label{eq:diffusionP}
\end{equation}

An initial qualitative observation of the line profiles, shown in Figure \ref{fig:SThMProfiles}, already allows for important conclusions. The Ge and Al sections of the wire exhibit very different characteristics. In the Ge segment, the temperature profile is approximately parabolic, as expected for a diffusive 1D system with uniform Joule dissipation. A heat source term in equation\ref{eq:diffusion} generally leads to a concave temperature profile. In contrast, the profile in the Al sections is convex, indicative of a situation which is dominated by heat dissipation and loss to the substrate. Here, the approximately exponential decay is expected for a uniform wire without any local Joule dissipation. The Al-Ge interface marks the Joule profiles by a sudden change of slope. When extracting the points of maximum temperature gradient for the Joule signals, their relative distances coincides for all measurements exactly with the length of the Ge segment. Also, after the deflection points there is a transition length of about 40-50\,nm in the Al-segments. This area corresponds to the phonon-electron thermalization length that is blurred by a combination of spatial resolution and parallel heat transport in the oxide shell (the Figure is in the supplemental information).

\subsection{Analysis of the aluminium leads}
\label{sec:Al}
\begin{figure}[tb]
	\centering
	\includegraphics[width=1.0\columnwidth]{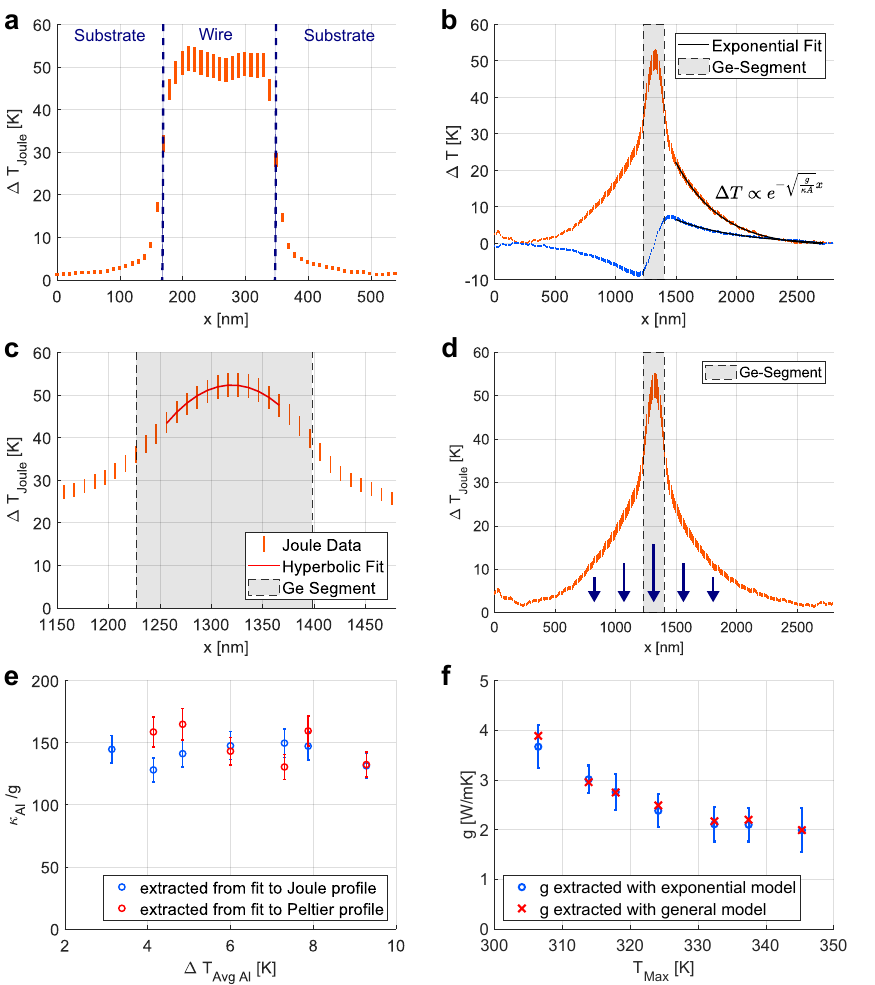}
	\caption{Visualization of the parameter extraction. a) The horizontal profile of the temperature change with respect to RT that caused by Joule heating $\Delta T_{\rm Joule}$. The discontinuity in the temperature profile when going from wire to substrate justifies an analysis in one dimension. b) The exponential fit in the Al-segment allows to extract the ratio between the thermal conductivity $\kappa$ and the substrate coupling constant $g$. The thermal data is marked with the corresponding uncertainty. c) The thermal conductivity in the Ge-segment is extracted from a hyperbolic fit to the measured temperature change caused by Joule heating in the nanowire. d) Finally, the substrate coupling constant is calculated via the observation, that all heat dissipated in the wire is lost into the substrate. e) Shows the values for $\frac{\kappa_{\rm Al}}{g}$ extracted from the exponential fit for the Peltier and Joule profile for different operating currents. f) presents the values for the thermal substrate coupling $g$ extracted from the more general model as in d) (in red) and the values extracted from the exponential fit (in blue).}
	\label{fig:fits}
\end{figure}
We turn first to the sections of the line profiles, in which the core is single crystalline Al. In these sections, the heat diffusion equation has no source terms and therefore reduces to  
\begin{equation}
\frac{\partial^2}{\partial x^2} \Delta T_{\rm j,p}(x)= \frac{g}{\kappa_{\rm Al}A}\cdot \Delta T_{\rm j,p}(x).
\end{equation}
The indices j and p denote Joule and Peltier terms, respectively. As the temperature profile reaches room temperature before reaching the electrodes, the analytic solution to this differential equation is given by
\begin{equation}
\label{eq:exponetial}
\Delta T_{\rm j,p}(x-x_0) = T_0 \cdot \exp{-\sqrt{\frac{g}{\kappa_{\rm Al} A}}\cdot (x-x_0)},
\end{equation}
where $T_0$ is the temperature at point $x_0$. Thus, the exponential fit to the temperature data reveals the relation between the substrate coupling constant $g$ and the thermal conductivity in Al $\kappa_{\rm Al}$. The exponential fit is applied to both the Peltier and Joule profiles under different operating currents. It is shown in Figure \ref{fig:fits}b. The extracted ratios for $\kappa_{\rm Al}/g$ are similar for all measurements, as can be seen in \ref{fig:fits}e. The values for $\kappa_{\rm Al}/g$ suggest, that the heat loss to the substrate is about $150$ times smaller than the heat conduction to the sides in each infinitesimal element d$x$. The assumption of the uniformity of the temperature distribution over a cross-section is, thereby, in retrospect justified by this result. Furthermore, the high consistency of the values derived from the $1\times f_{\rm mod}$ and $2\times f_{\rm mod}$-measurements is an experimental confirmation for negligible Joule heating or Peltier heating/cooling in this section.

The analysis is pushed further by deriving $\kappa_{\rm Al}$ and $g$ independently. It is observed, that the wire temperature of the Joule profiles decays to room temperature before reaching the electrodes. In other words, the entire heat that is generated over electric power dissipation in the wire ($P_{\rm dis} = I^2 \rho L_{\rm Ge} / A_{\rm core}$) is equal to the heat lost to the substrate. By making use of a finite element integral over all the measuring points on the nanowire, the substrate coupling is then the only unknown quantity. The conservation of heat requires
\begin{equation}
\begin{split}
P_{\rm dis} &= \int\limits_{\rm NW} g\cdot \Delta T_{\rm Joule}(x) {\rm d}x  \\
&\approx g\cdot\Delta x \cdot \sum\limits_{\rm NW} \Delta T_{\rm Joule}(x_i),
\end{split}
\end{equation}
where $\Delta x \approx 10$\,nm is the spacing between adjacent  measurement points $x_i$, $g$ the substrate coupling, $T$ the sample temperature and $\Delta T_{\rm Joule}(x_i)$ the measured temperature increase of the finite element $x_i$ caused by Joule heating with respect to ambient temperature. With this approach we extracted values for $g$ of $2.1\pm0.3$, $2.1\pm0.3$ and $2.0\pm0.4$\,W/(mK) for the experiments using 23, 24, and 26\,$\mu$A, respectively.

We note that one reaches the same result by considering continuity of heat flux at the Ge-Al interface in addition to the 1D heat diffusion equation. The heat transported into the Al segment is given by $\dot{Q}_{\rm Al} = \grad T(x_0) \cdot \kappa_{\rm Al} A_{\rm wire}$. The temperature gradient is then calculated from the exponential fit and extrapolated to the Al-Ge interface. Energy conservation implies:
\begin{equation}
P_{\rm dis} = 2\dot{Q}_{\rm Al} + g\cdot \Delta T_{\rm avg\, Ge} \cdot L_{\rm Ge},
\end{equation}
where $L_{\rm Ge}$ is the length of the Ge segment and $T_{\rm avg Ge}$ is the average temperature along the Ge segment. Solving this latter equation based on the exponential fit leads to the same results as previously, this is shown in blue on Figure \ref{fig:fits}f. The good agreement with the result based on the more general result of energy conservation strengthens the diffusion equation based analysis further. As argued above, the value of $g$ is dominated by the thermal interface between the oxide shell and the substrate. The contact width is $(56 \pm 5)$\,nm (see Appendix). Therefore, we can calculate a value per unit area of approximately $3.5\cdot 10^7$\,W/(Km$^2$). This is a typical value for a thermal interface between dielectrics.

Next, we turn to the thermal conductivity of the Al segment. Using the extracted values of $g$ and the ratio $\kappa / g$, we arrive at a thermal conductivity for the Al segments. To interpret, we need to consider that the value from the fit is an average over the cross section of both the Al core and the Al$_2$O$_3$ shell which are weighed by the respective thermal conductivities and obtain for the core $\kappa_{\rm Al} \approx 150$\,W/(mK). We use the measured cross-sectional areas from the STEM image and the tabulated values for Al$_2$O$_3$. To compare with expected values we apply the Wiedemann Franz law, which results in an electrical resistivity of $\rho_{\rm Al} \approx 50\cdot 10^{-9} \, \Omega$m about twice as high than theoretical predictions for bulk c-Al \cite{Gall2016,Jain2016} which is expected due to increased surface scattering in nanostructures. Indeed, this value is three times lower than previously determined for pure c-Al wires \cite{Brunbauer_2016}. However, in an SThM scan of such an operating c-Al wire a dominant local heat source is identified in the center (see the temperature profiles in the supplemental information). This may be expected according to the wire fabrication and indicates a grain boundary. Therefore, the expected values for the electrical conductivity of the c-Al segments without any grain boundary are significantly higher and in good agreement with our experimental value.  This result is shows that the influence of grain boundary scattering on the electrical resistivity exceeds the impact of surface scattering in this device. This is typical for polycrystalline nanowires of these dimensions\cite{Durkan2000}. Finally, we would like to mention the importance of taking into account the contribution of an oxide shell to the heat transport in a nanowire.

\subsection{Analysis of the germanium section}
Next, a value for the thermal conductivity in the Ge-segments is extracted. We note, that the conduction of electrons is limited by the area of the core, whereas the conduction of phonons occurs in both, the Ge and the Al$_2$O$_3$-shell. Equation \ref{eq:diffusionJ} becomes
\begin{equation}
\label{eq:GeDGL}
\frac{\partial^2}{\partial x^2}\Delta T_{\rm j} + \frac{g}{\kappa_{\rm Ge}A_{\rm wire}}\Delta T_{\rm j}= - \frac{I^2 \rho_{\rm Ge}}{A_{\rm core}} \frac{1}{A_{\rm wire}\kappa_{\rm Ge}}.
\end{equation}
Here, $\kappa_{\rm Ge}$ is an average value over core and shell in the Ge segment.
This second order differential equation is analytically solved by
\begin{equation}
\begin{split}
\label{eq:GeSolution}
\Delta T_{\rm j}(x) = \left( \frac{\Delta T_1 + \Delta T_2}{2}- \frac{q}{g} \right) \frac{\cosh (m(x-x_c))}{ \cosh (mL/2)} \\ + \frac{\Delta T_2 - \Delta T_1}{2} \frac{\sinh (m(x-x_c))}{ \sinh (mL/2)} + \frac{q}{g}
\end{split}
\end{equation}
where $x_c$ is the center on the nanowire, $\Delta T_{1,2}$ is the temperature at the boundaries of the fit, the Joule dissipation per unit length $q = I^2 \rho_{\rm Ge} / A_{\rm core}$, and $m = \sqrt{g/k_{\rm Ge} A_{\rm wire}}$. 
This function is fitted to the temperature profile in the Ge-segment. In the Casimir result, the mean free path of phonons is limited by the diameter of the nanowire, although in this core-shell system the mean free path may be even smaller than that. The heat diffusion equation, equation\,\ref{eq:diffusion}, is valid for problems on a larger length scale than the mean free path. Therefore, the points lying closer to the thermal interface than the diameter are excluded for the fitting. Furthermore, to reduce uncertainty of the fit, we fix $g$ from the analysis above and use only $\kappa_{\rm Ge}$ and the boundary conditions as a fitting parameter in equation \ref{eq:GeSolution}.
The fit is shown in Figure \ref{fig:fits}c, and the extracted value for the thermal conductivity $\kappa_{\rm Ge}$ is $16.9\pm2.5$\,W/m/K at a current of 26\,$\mu$A.

It is interesting to note, that in this case of relatively small values of $m$, the temperature profile looks parabolic. For $g=0$, supposing that the temperature profile is dominated by Joule heating and heat evacuation towards the electrodes, the solution to equation \ref{eq:GeDGL} would indeed be a parabola. Such a fit seems to match the data well. However, the extracted value is given by $\kappa_{\rm Ge} = 23.7 \pm 3.3$\,W/m/K higher by 40\% and we conclude that the hyperbolic model should not be simplified.

To interpret the result, we consider that the extracted value for $\kappa_{\rm Ge}$ comprises the contributions of both oxide shell and Ge core. The area of the shell is double the area of the core, however, the exact values are measured on the STEM image. Both Al$_2$O$_3$ and Ge crystals have a larger conductivity in bulk, of about 30 and 50 to 60\,W/m/K, respectively. Both values, however, are expected to be significantly reduced, due to the boundary scattering of the phonons at reduced dimensions\cite{Mingo2003} and the measured value is consistent with expectations.  

\subsection{Extraction of the Peltier coefficient and thermoelectric properties}
\begin{figure}[tb]
	\centering
	\includegraphics[width=1.0\columnwidth]{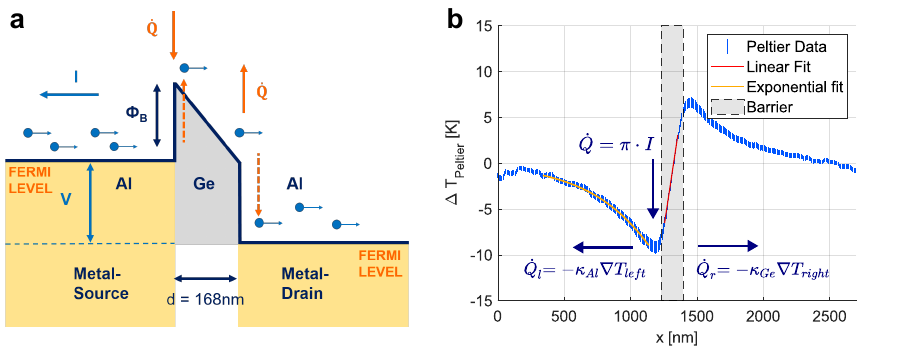}
	\caption[A floating figure]{Illustration of the Peltier effect a) The Fermi energy of electrons in a metal is lower than the average transport energy in a semiconductor. When electrons pass from a metal to a semiconductor they absorb energy from the lattice which creates a local heat sink. On the other side, when electrons pass from the semiconductor to the metal the higher energy electrons thermalize with the lattice which creates a local heat source.
	b) The Peltier coefficient is extracted by considering, that the injected heat is equal to the heat flux to the left and the heat flux to the right. The temperature gradients are calculated by help of an exponential fit in the Al-segment and a linear fit in the Ge-barrier.}
	\label{fig:peltier}
\end{figure}

At this length scale, the distinction between the Peltier effect and thermionic emission becomes blurry. Whereas the Peltier effect exists also in bulk materials, the thermionic effect is understood in a microscopic regime around an energy barrier present at the interface of a heterostructure\cite{thermoelec_book}. This region is marked by the thermalization length of the non-equilibrium charge carriers with the lattice. In this analysis, an effective Peltier coefficient is assigned to the thermionic cooling and heating by charge carriers transported over a potential barrier. The local Peltier coefficient is associated with a length scale $\lambda$  through $\pi (x) = (\Pi / \lambda) \exp (-(x-x_0)/\lambda)$. On the scale of $\lambda$ the non-equilibrium charge carriers equilibrate with the lattice (i.e. the phonon system) after passing the the metal-semiconductor interface. Indeed, we observe this distance to be larger than the scattering length of charge carriers for electron-phonon scattering of about 19 to 25\,nm\cite{haynes2014crc, Kanter1970}. 

The extraction of the Peltier coefficient is based on the conservation of heat flux at steady state. In the Al wires and near the Al-Ge interfaces, there is either a heat sink or a heat source caused by the Peltier effect $\dot{Q}_{\rm GeAl}=\Pi\cdot I$ or $\dot{Q}_{\rm AlGe}=-\Pi\cdot I$, respectively, where $\dot{Q}_{\rm AlGe}>0$ and $\dot{Q}_{\rm GeAl}<0$, and $\Pi$ is the Peltier coefficient. An illustration is seen in Figure \ref{fig:peltier}b. The injected heat $\dot{Q}$ from these regions is then equal to the sum of the heat flux towards the Al-electrodes and the heat flux over the Ge-segment to the other heat source/sink. (We can again neglect on the short length scale $\lambda$, as discussed above.) By use of Fourier's law the relation is then written as
\begin{equation}
\dot{Q}= (-\kappa_{\rm Al}\nabla T_{\rm Al} - \kappa_{\rm Ge}\nabla T_{\rm Ge})\cdot A_{\rm wire}.
\label{eq:injectedHeat}
\end{equation}
In order to take into account the above described microscopic extension of the Peltier coefficient we consider the temperature profile beyond the length scale away from the Al-Ge interface. Then we can treat the temperature profiles as solutions to equation \ref{eq:diffusionP} in sections without source terms.
The temperature gradient in the Al-segment is the derivative with respect to the position $x$ of the exponential fit to the Peltier profile evaluated at the junction as $\nabla T_{\rm Al} = \sqrt{\frac{g}{\kappa A}} \cdot\Delta T_{\rm AlGe}$, see equation\,\ref{eq:exponetial}. By extrapolating the exponential fit we conceptually concentrate the extended heat source in a point at the AlGe interface. The temperature gradient in the Ge-segment is obtained from a linear fit to the data between heat source and heat sink. The linearity of the temperature profile within the Ge section is a result from the strong temperature gradient and the $\kappa / g$ ratio which justifies neglecting substrate loss $g$ in this region. 
The effective Peltier coefficient is then calculated as $\Pi=267\pm25$\,mW/A, resulting in a Seebeck coefficient of $S=790\pm95\,\mu$V/K for the temperature at the Al-Ge interface. 

It is interesting to relate the effective Peltier coefficient to the Schottky barrier height to test the validity of the derived value. The cooling power by thermionic emission over a barrier is given by the total current times the average energy of the carriers over the barrier
\begin{equation}
\dot{Q}=I\cdot\bigg(\phi_C+\frac{2k_BT}{e}\bigg).
\end{equation}
Hence, the barrier height may be calculated from our measurements. We find $\phi_{\rm c}=325\pm45$\,meV, which is in agreement with previously determined barrier heights from gating experiments of similar devices\cite{Kral2015}. Independent of metal type and doping concentrations, metal-germanium junctions form Schottky contacts and exhibit very strong Fermi-level pinning close to the valence band \cite{THANAILAKIS19731383, Dimoulas2006}.
Finally, the thermoelectric figure of merit is calculated for the scan taken at a current of $I=26\,\mu$A to be $ZT = 0.020\pm0.005$. This finalizes the thermoelectric characterization of the Al-Ge heterostructure nanowire with a segment length of 168\,nm from a single measurement.

Lastly, we have a closer look at the Ge-Al interface regions. We acknowledge that extracting quantitative results is hampered by the oxide shell, which does not comprise an interface. However, important observations can nevertheless be made. The conjecture that the Peltier source term has an exponential spatial distribution has testable consequences. As a result of the distributed source term (equation \ref{eq:petcoefficient}), the point of maximum temperature should be shifted with respect to the Al-Ge interface. Using the differential equation with this source term and the material parameters extracted above, we calculate that the maximum temperature should be about 46\,nm separated from the interface using an equilibration length of 22\,nm. This shift is a good estimation with the position of maximum temperature extracted from the experimental temperature profile of $58 \pm 10$\,nm. 

\section{Conclusion}
This study demonstrates the capability to use thermal imaging by SThM for extracting relevant information on material properties and the dynamics of nanoscale devices and structures. Most notably, we were able to obtain the thermal and thermoelectric properties of a model heterostructure from a single measurement. By quantifying the thermal contact resistance between sample and substrate, its effect can be quantified and, in contrast to other techniques, does no longer hamper the measurements. The values are confirmed by independent measurements, such as the derivation of the substrate coupling with two different models and the comparison of the barrier height investigated thermally with previous electrical transport measurements. It appears, there is currently no other method available to extract this set of information from a sample of this size. For thermoelectric applications, the performance of short segments is interesting, and the results create a link between device design and materials properties. 

\section{Acknowledgements}
This project has received funding from the European Union’s Horizon 2020 research and innovation programme under Grants 482
No. 766853 (EFINED) and No. 767187 (QuIET).

Continuous support from Anna Fontcuberta i Morral, Ilario Zardo, 
Marta De Luca, Kirsten Moselund, Heike Riel and Steffen Reidt is gratefully acknowledged.

\bibliography{bibliography.bib}


\end{document}


\title{Supplementary Information: Spatially Resolved Thermoelectric Effects in operando Semiconductor-Metal Nanowire Heterostructures}

\author{Nadine G\"achter}
    \email[Correspondence email address: ]{nadine.gaechter@unibas.ch}\affiliation{IBM Research - Zurich, 8803 R\"uschlikon, Switzerland}

\author{Fabian K\"onemann}
    \affiliation{IBM Research - Zurich, 8803 R\"uschlikon, Switzerland}

\author{Masiar Sistani}
    \affiliation{Institute of Solid State Electronics - TU Wien, 1040 Vienna, Austria}

\author{Maximilian G. Bartmann}
    \affiliation{Institute of Solid State Electronics - TU Wien, 1040 Vienna, Austria}

\author{Marilyne Sousa}
    \affiliation{IBM Research - Zurich, 8803 R\"uschlikon, Switzerland}

\author{Philipp Staudinger}
    \affiliation{IBM Research - Zurich, 8803 R\"uschlikon, Switzerland}

\author{Alois Lugstein}
    \affiliation{Institute of Solid State Electronics - TU Wien, 1040 Vienna, Austria}

\author{Bernd Gotsmann}
    \email[Correspondence email address: ]{bgo@zurich.ibm.com}
    \affiliation{IBM Research - Zurich, 8803 R\"uschlikon, Switzerland}

\date{\today} 

\maketitle
\section{Device Fabrication}
The NWs are fabricated by thermally induced substitution of gold assisted vapor-liquid-solid (VLS) grown Ge-NWs by Al. At first, <111> oriented Ge-NWs are grown heteroepitaxially on silicon substrates in a low pressure chemical vapor deposition system by use of a gold-assisted VLS process. The NWs are then coated with 15\,nm high-\textit{k} Al$_2$O$_3$ by atomic layer deposition (ALD) and drop-cast on a highly p-doped Si substrate with a 100\,nm thick dielectric layer of SiO$_2$. Afterwards, the Al-contacts are formed with electron-beam lithography, sputter deposition and lift-off techniques in preparation for the subsequent thermal exchange reaction. Lastly, in order to form the single crystalline Al-Ge-Al heterostructure, the Ge-NWs are thermally annealed at 623\,K. Under these conditions, the diffusion constants of Ge and Al are $10^{12}$ times higher in Al than in Ge\cite{Elhajouri2019}. This means, that the Ge can easily diffuse into the Al-contact pats, whereas the Al-atoms are efficiently supplied via fast self-diffusion to take over the released lattice sites. The successive substitution of Ge-atoms with Al has been monitored \textit{in-situ} in both, scanning electron microscopy (SEM) and scanning transmission electron microscopy (STEM) studies. They revealed atomically sharp interfaces between the Al and Ge during the anneal. Thus, the length $L_{\rm Ge}$ of the Ge-segment is tuned by varying the annealing time. For sufficiently long baking times the Ge diffuses completely out of the wire into the contact pads. It leaves behind single crystalline Al (c-Al) NWs with a single grain boundary in the center. Furthermore, the Ge-segments are integrated in a back-gated field effect transistor (FET) by construction. The Al-contacts are naturally self-aligned with the wire. More details are found in references \cite{Kral2015,Elhajouri2019,Brunbauer_2016}.

\section{NW properties}
Figure 1 b-d in the paper show a schematic, a STEM image of the cross-section and a SEM of the nanowire heterostructure studied here. More particularly it consists of a single-crystalline Ge-segment of 168\,nm length and 37\,nm diameter, as determined using scanning transmission electron microscopy (STEM). The Ge-segment is contacted by two self-aligned c-Al nanowires of the same diameter and around 1.7\,$\mu$m length. An aluminium oxide shell (Al$_2$O$_3$) of 15\,nm thickness deposited by atomic layer deposition (ALD) surrounds the Al-Ge-Al core. The interfaces between the c-Al and Ge are abrupt to the atomic level. The Al-sections of the nanowire are contacted using Al pads patterned with electron-beam lithography on a Si/SiO$_2$-substrate. 

The electrical properties of pure c-Al nanowires fabricated using the same process revealed a conductivity of $\sigma = (7.6 \pm 1.5) \cdot 10^6 \, (\Omega$m$)^{-1}$ for Al\cite{Brunbauer_2016}. The Al-Ge-Al heterostructures with atomically abrupt interfaces show the non-linear current-voltage (IV) relationship of two back-to-back Schottky diodes in series for Ge-segments of length $L_{\rm Ge}>45$\,nm with an overall resistance  proportional to the Ge-segment length\cite{sistani2017}. From gating experiments it is further concluded that the Ge segments act like p-type semiconductors. Independent of metal type and doping concentrations, metal-germanium junctions form Schottky contacts, exhibiting very strong Fermi-level pinning close to the valence band \cite{THANAILAKIS19731383, Dimoulas2006}.

One issue regarding the thermal characterization of current-carrying Al-Ge-Al nanowires are traps for charge carriers located at the Ge/Ge-oxide interface. In this regard, the protective Al$_2$O$_3$-shell ensures reliable and reproducible measurements by avoiding any influence of adsorbates rather than eliminating charge trapping due to dangling bonds at the Ge/Ge-oxide interface. These traps lead to hysteretic current voltage relationships. However, the hysteresis decreases with increasing current range, drive speed and operating time. It was found that for currents larger than 21\,$\mu$A, the hysteresis is reduced and the current-voltage graph sufficiently linear to assume a constant resistance in the thermal analysis. Consequently, we estimate a contribution to the systematic error to the extracted temperature values for lower currents shown below due to these charging effects.

\section{Imaging and Geometric Characterization of the Wire}

\begin{figure}[tb]
	{\includegraphics[width=1.0\columnwidth]{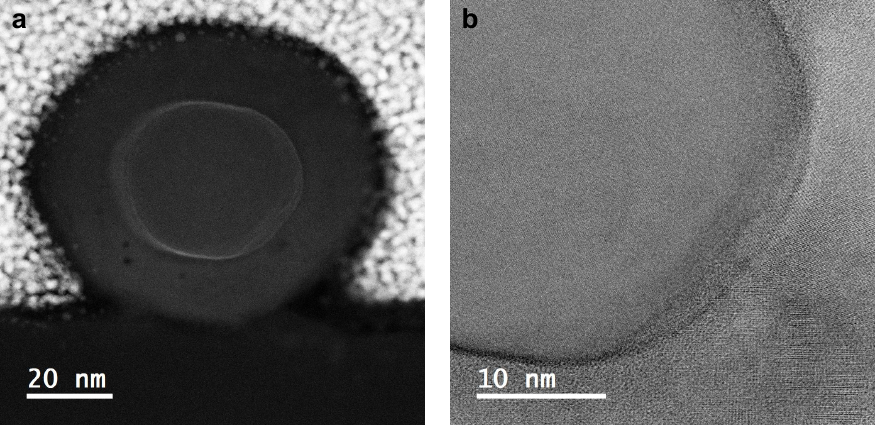}}
	\caption{DF-STEM scans taken after the thermal measurements. The oxide Aluminium Oxide layer shows a certain degree of crystalinity which enhances the thermal conductivity.}
	\label{fig:STEM}
\end{figure}

After the thermal scans, a scanning transmission electron microscopy analysis of an Al segment is carried out with a double spherical aberration-corrected JEOL JEM- ARM200F operated at 200 kV. The Al segment was prepared by focused ion beam, using a dual beam FIB Helios NanoLab 660 from FEI. The DF STEM images of the cross-section are shown in figure \ref{fig:STEM} and reflect the round shape. A sharp interface separates the core and the oxide shell with respective areas of $A_{\rm core} = 1075\pm 80 $\,nm$^2$ and $A_{\rm shell} = 2450\pm230 $\,nm$^2$. The Al-core shows some poly-crystallinity, however, the loss of crystalline order probably occurred during sample preparation in the FIB. Previous more extensive studies suggest crystalline order. Furthermore, the Al$_2$O$_3$-shell is observed to be crystalline in the Fourier image of the cross-section (not shown here). 

\begin{figure*}[tb]
	\centering
	\includegraphics[width=2.0\columnwidth]{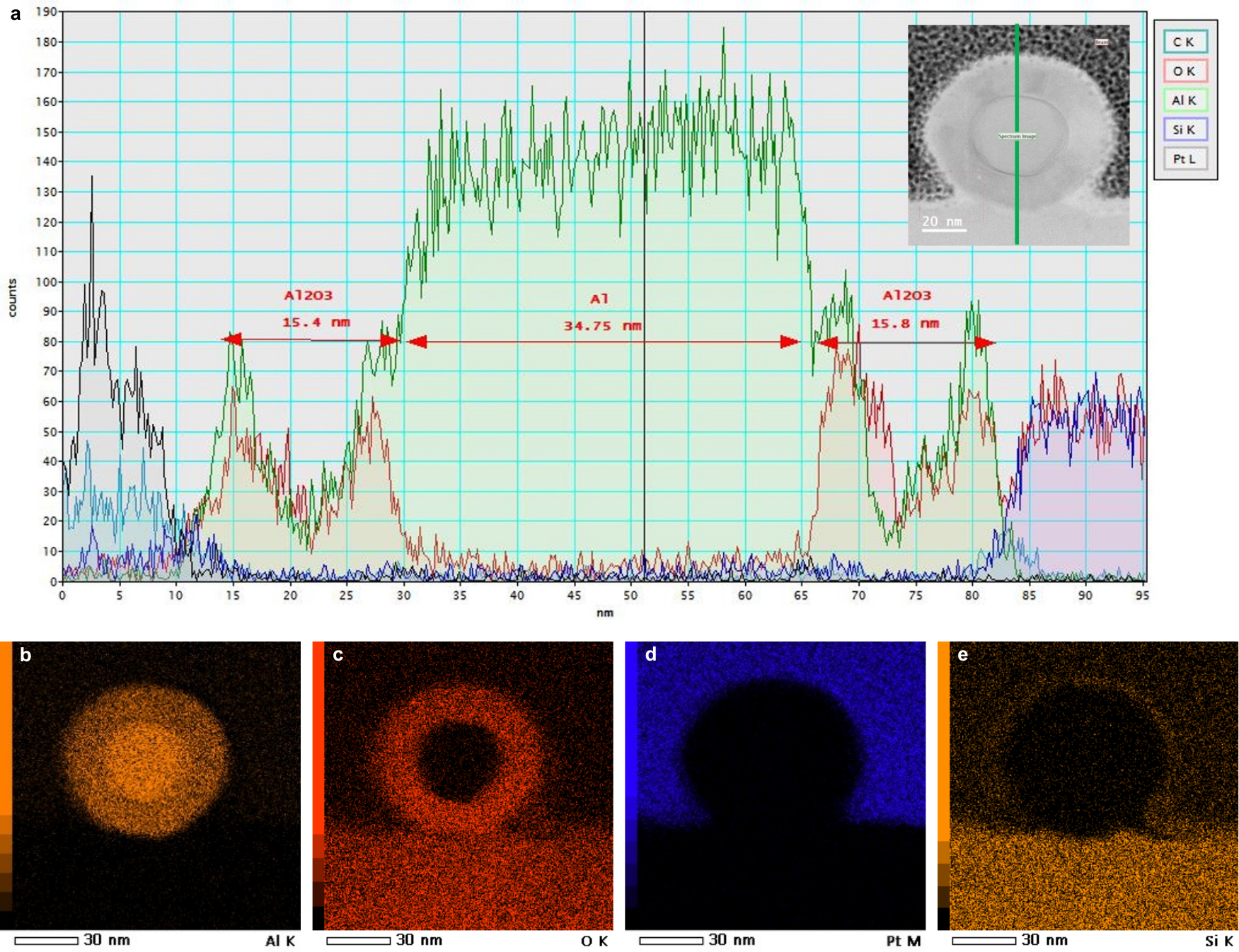}
	\caption{Chemical Analysis (EDS) of the nanowire cross-section. a) Along the a line through the center of the wire's cross-section, see the green line on the STEM image in the inset. b), c), d), e) show the number of counts at the energy of the Al, O, Pt and Si respectively. They demonstrate the high purity of the different materials. The platinum was deposited after the thermal measurements, to protect the wire before the FIB cut.}
	\label{fig:eds}
\end{figure*}

Energy Dispersive X-ray Spectroscopy (EDS) chemical maps reveal a well defined Al-core and Al$_2$O$_3$ shell with a sharp transition between the two. No residual Ge is detected beyond the detection limit which does not resolve doping concentrations. They are shown in figure \ref{fig:eds}. To conclude, all geometric properties of the NW are measured and summarized in table \ref{tab:geometry}. 

\begin{figure}[tb]
	{\includegraphics[width=0.8\columnwidth]{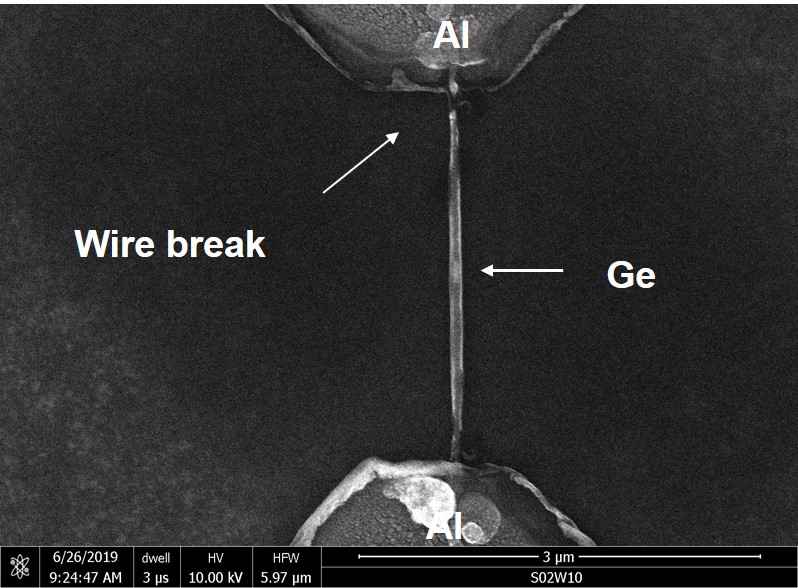}}
	\caption{The SEM image after the thermal scans when the wire was no longer conducting. The breaking point lies close to the electrodes, even though the thermal scans revealed only little heating in these areas.}
	\label{fig:sem_after}
\end{figure}

Finally, figure \ref{fig:sem_after} shows the SEM-image taken in the FIB, when the wire was no longer conducting. The location at which the wire broke, lies close to the upper Al-electrode. In most of the experiments the wires break in a similar spot, even though the heating is observed closer to the Ge-segment. The wire is slightly thinner towards the electrodes due to the chemical etching of the oxide shell before the deposition of the electrodes.


\begin{table}[tbh]
	\centering
	\begin{tabular}{|c|l|} 
		\hline
		$L_{\rm wire}$ & $(2475\pm5)$\,nm\\ \hline
		$L_{\rm Ge}$ & $(168\pm1)$\,nm \\ \hline
		$r_{\rm core}$ & $(18.5\pm2)$\,nm \\ \hline
		$d_{\rm oxide}$ & $(15\pm2)$\,nm \\ \hline
		$d_{\rm touch}$ & $(56\pm5)$\,nm \\ \hline
		$A_{\rm core}$ & $(1075\pm 80) $\,nm$^2$ \\ \hline
		$A_{\rm shell}$ & $(2450\pm230 $\,nm$^2$ \\ \hline
		$A_{\rm wire}$ & $(0.0035\pm0.0003)\, \mu $m$^2$ \\ \hline
	\end{tabular}
	\caption{Geometric properties of the NW obtained from SEM and STEM images: $L_{\rm Ge}$ is the length of the Ge segment, $r_{\rm core}$ is the radius of the conducting core, $d_{\rm oxide}$ is the thickness of the surrounding Al$_2$O$_3$ shell, $d_{\rm touch}$ is the length perpendicular to the wire that is in contact with the substrate. $A_{\rm core}$, $A_{\rm shell}$ and $A_{\rm wire}$ are the cross-sections of the conducting inner core, the passivating Al$_2$O$_3$ shell and the entire wire respectively. The error estimation for the areas follows Gaussian error propagation.}
	\label{tab:geometry}
\end{table}


\section{Electrical Characterization of the nanowire}

\begin{figure*}
	\centering
	\includegraphics[width=2.0\columnwidth]{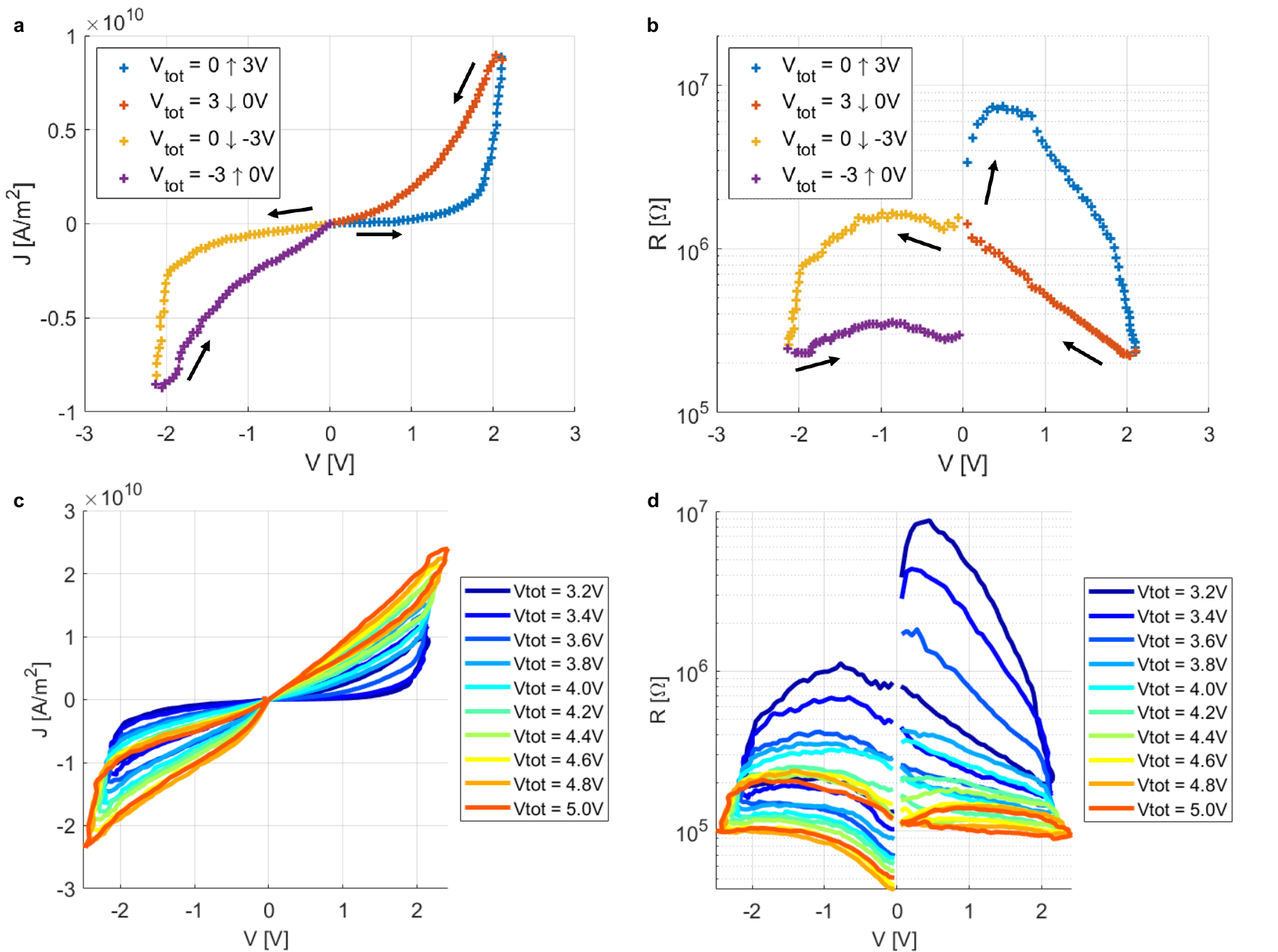} 
	\caption[A floating figure]{Characterization of the electrical behavior of the nanowire. The data in a) and b) is measured just below a current density $J$ is that is high enough to lead to a detectable change in the temperature for the wire. a) is the current density is represented as a function of the voltage drop over the device, b) the same information shown with the device resistance as a function of the voltage drop over the NW. c) and d) show the I-V and R-V curves measured immediately after each SthM scan up to the same voltage bias. The total applied voltage $v_{\rm tot}$ drops additionally over a series resistance of $R_{\rm S}=100$\,k$\Omega$.}
	\label{fig:ivbeforeheating} 
\end{figure*}

This section provides some overview on the electrical behavior of the NW under measuring conditions for the thermal scans. Figure \ref{fig:ivbeforeheating} displays the I-V relation that is taken just below the threshold voltage at which any heating of the wire is detected. The I-V has two very striking features: First, a pronounced hysteresis appears between ramping the current up and down. Second, the electrical resistance depends highly on the applied voltage. Both observations are well visible on the resistance-voltage (R-V) plot in the logarithmic scale depicted next to the I-V. It reveals a change in resistance of two orders of magnitude. A heating of the wire is observed only if the applied electric field is higher than the electrical breakdown field in pure Ge wires that has been previously observed at $E=1.25\cdot10^5$\,V/cm \cite{Lugstein_2013}. For a Ge-segment of length $L_{\rm Ge} = 168$\,nm this corresponds to an applied voltage bias of $V_{\rm bias}=2.1$\,V.

The wire is integrated in an electrical circuit including a series resistor of similar resistance $R_{\rm S} = 100$\,k$\Omega$ that protects from current spikes. The real time data acquisition system is is also used to measure the IV after each scan. The IV is conducted in the DC-mode with 200 steps distributed over the AC-modulation. Each point is averaged over $20$\,ms.

\begin{figure*}
	\centering
	\includegraphics[width=0.8\columnwidth]{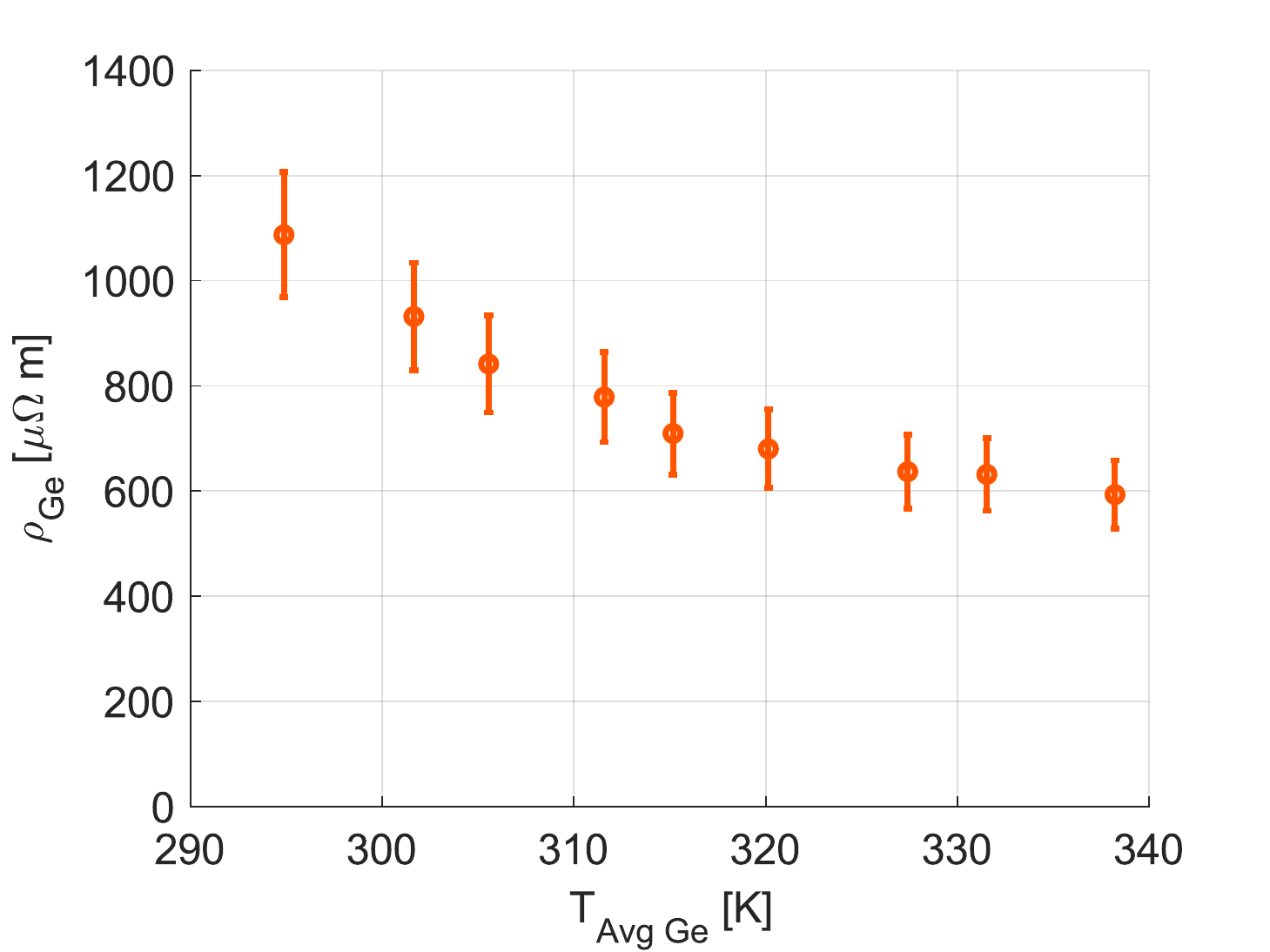} 
	\caption[A floating figure]{The electrical resistivity $\rho_{\rm Ge}$ is shown as a function of the average temperature in the Ge segment.}
	\label{fig:rhoge} 
\end{figure*}

Figure \ref{fig:ivbeforeheating} shows a plot of all I-V and R-V curves going up to the maximum voltage bias applied in the respective thermal measurements. The electric field in the Ge-segment is between $E \approx V_{\rm device}/L_{\rm Ge} = 1.27-1.44\cdot10^5$\,V/cm. Each curves is recorded after one hour of operation time and, therefore, charging of the oxide shell during the thermal scan. The wires have some seconds to cool down after the scan. The very different behavior of the IVs for different maximally applied voltages is, therefore, not caused by changes in the device temperature which should be the same for equal current density. Hence, it might be attributed to the charging effect in the oxide shell. Moreover, the amount of charging depends upon whether the NW is operated under AC- or DC-current. It is observed, that this local gating phenomenon changes the resistance of the device by orders of magnitude for only small changes of the experimental condition. However, the amount of charging tends to a stationary state when the IV is run quickly several times. Therefore, the hysteresis is expected to disappear sufficiently during the scans in the SThM. A difficulty for investigating thermoelectric effects is the simultaneous presence of Joule heating, which, on the other hand, creates a positive feedback phenomenon by decreasing the resistance during heating. It is counteracted only by an increased charge carrier scattering at higher temperature.

An increase in drain current with the gate voltage is linked to the trapping of negative surface charges in the interband levels due to surface states and bulk impurities. An applied negative gate voltage provokes the accumulation of holes that continuously neutralize the trapped electrons\cite{sistani2017}. A similar p-type charging has been shown to appear in undoped Si-NWs. In the Si-system, changing the passivation layer from Al-oxide to Si-oxide turns the device into a n-type FET, being OFF for negative gate voltage. Therefore, surface charges originating from interface defects and dangling bonds can generate doping effects in NWs\cite{Winkler2015}. The use of a high quality thermal oxide reduces the number of charge traps and thereby also the hysteresis that occurs when the gate voltage is sweeped from negative to positive values and vice versa\cite{sistani2017}. Under a given gate voltage, the neutralization of holes is a rather slow process. It reaches a stationary state after approximately 20 minutes of operation. Such a long time-span is owed to a kinetic limitation by either a diffusion or tunnel barrier in form of a GeO$_x$ layer at the interface between the Ge and the Al$_2$O$_3$. Indeed, such a layer is expected to form during ALD-deposition of the Al$_2$O$_3$ layer. It acts as a local gate and provides a large number of trapping states\cite{Zhang2015}\cite{staudinger2018}. To conclude, the nature of the passivation layer highly influences the reliability and reproducible performance of the device under the above described charging mechanism.

The electrical resistivity in the Ge-segment is determined as follows. It is calculated from the electrical resistance measured in the I-V curves which are taken after each scan. The resistance at the maximal voltage in the I-V is the same as the AC-voltage applied during the scan closest to the conditions during the respective thermal measurement. The contribution of the Al-wire and contacts is subtracted from the overall device resistance. The values are represented as a function of the average temperature in the Ge-segment in figure \ref{fig:rhoge}. The electrical resistivity depends strongly on the temperature.

\section{SThM operation and analysis details}
The images of the scans are generated and treated with the help of the open-source software Gwyddion. Apart from the straight forward calculation of the temperature field using equation 1 in the paper, some further data processing is conducted: Firstly, under the observation that the trace and retrace are nearly identical, they are averaged for noise reduction. Secondly, a low-pass scaling is applied to compensate for the signal attenuation that is due to a delay in temperature response of the cantilever. Thirdly, a DC-signal offset is manually corrected to eliminate some remaining non-uniformity in the temperature field which is caused by the artifacts in the DC-field. It is hard to measure the temperature offset of the sensor exactly due to small changes in DC-signal when approaching the tip. However, by eye it is easy to see when the artifacts at the edge of the NW disappear. The procedure might be compared to aligning a microscope. The final heat maps are shown in figure 2.

\begin{figure}[tb] 
	\centering
    \includegraphics[width=1.0\columnwidth]{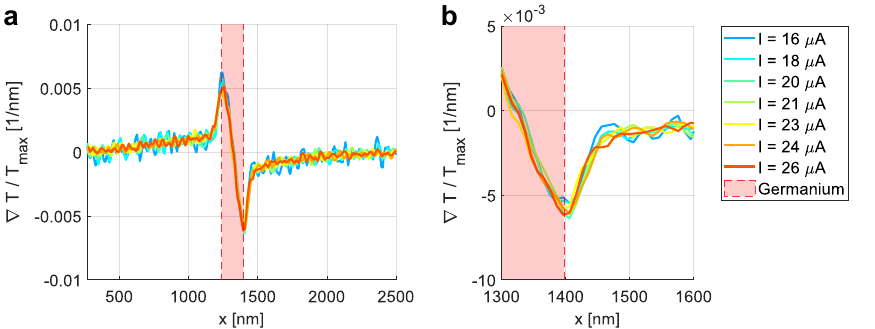}
	\caption{Determination of the location of the Ge segment is possible due to a discontinuity in the temperature gradient at the thermal interface between the Ge and the Al segments. a) The temperature gradient $\nabla T_{\rm Joule}$ as a function of the position $x$ along the wire for the different operating currents. The gradients are normalized by the maximum temperature in the center of the red shaded Ge segment. b) shows a zoom on the gradient at the Ge-Al interface.}
	\label{fig:gradient} 
\end{figure}

When turning the focus to the temperature profile around the Al-Ge interface, the location of the Ge segment on the temperature line has to be determined first. The temperature gradient of the Joule profile is plotted in figure \ref{fig:gradient}a). A pronounced maximum and
minimum peak marks the deflection points in the Joule profile. They result from a sudden change of the thermal and electrical conductivities between the Ge- and the Al-parts. Their distance of 169\,nm corresponds precisely to the expected length of the Ge-segment in a scan with 10\,nm resolution. The hereby identified location of the Ge-segment is then shaded in red on the profile lines and used as orientation in the following analysis. After the deflection point there is a transition length of about 40-50\,nm, that is well seen on the zoom-in figure \ref{fig:gradient}b). This area corresponds to the phonon-electron thermalization that is blurred by a combination of the spatial resolution and a parallel heat transport in the oxide shell.

In general, the temperature profiles show mostly symmetric behavior for the Joule and the Peltier signal. It reflects the good quality of the measurement data and the sample. However for the thermal measurements that were conducted at lower operating currents, the profile curves show a slight asymmetry. This is when the SThM method reaches its limitations due to the underlying assumption of a linear electrical behavior of the sample. This requirement is better fulfilled at higher currents and leads to a mixing of the $1f_{\rm mod}$ and $2f_{\rm mod}$ signal at lower currents.
\begin{figure}[tb]
	\centering
    \includegraphics[width=1.0\columnwidth]{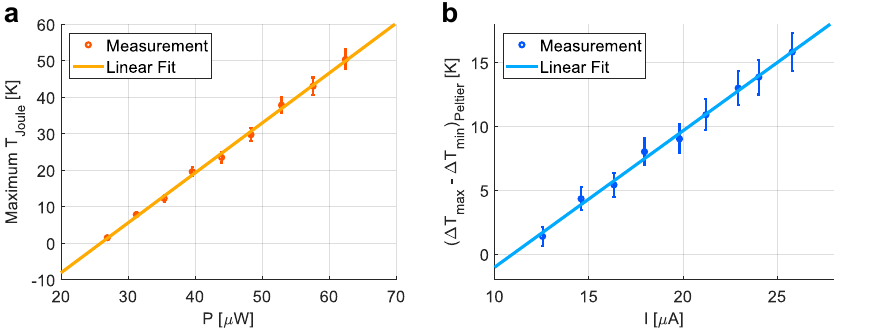} 
	\caption{In a), the maximum temperature increase caused by Joule heating is plotted as a function of the dissipated power $P$. In b), the temperature difference between the two poles in the Peltier scan as a function of the operating current. The errorbars reflect the uncertainties on the temperature measurement.}
	\label{fig:propI2I}
\end{figure}
Nonetheless, the comparison of the the maximum temperature in the Joule-scan with the dissipated power and the comparison between the maximum temperature difference in the Peltier-scan and the current reveal linear relationships. These results is expected under the the 1D diffusion equation 2 from the paper. They can be seen in figure \ref{fig:propI2I}.
The noise levels are quite low. No noise reduction is used at any stage in the lateral direction of the wire, to omit altering the shape of the profile. The expected error on the measured temperature is a combination of a constant absolute error and a more important relative contribution that depends on the sample temperature. The absolute error is determined by taking the root mean squared of the temperature on the substrate far from the wire where no signal is expected. The relative error the result of a multiplication with the temperature difference to the sensor according to equation 1 is estimated to be around $5\%$ of the sample temperature\cite{konemann2019}. The estimated uncertainty for each temperature measurement along the profile of the NW is then calculated as
\begin{equation}
\Delta \bigg(\Delta T _{\rm J/P}\bigg) = \frac{1.84  [\rm{K}]}{\sqrt{15}}+0.05\cdot\Delta T_{\rm J/P}
\label{eq:error}
\end{equation}
To conclude, the method allows to resolve the temperature profiles of the samples that are caused by the Joule and Peltier effects down to the thermalization lengths of the heat carriers. In order to extract further information by the most simple means, the analysis is divided in different sectors on the wire.

\section{Joule heating in pure Aluminium nanowires}

\begin{figure}
	\centering
	\includegraphics[width=1.0\columnwidth]{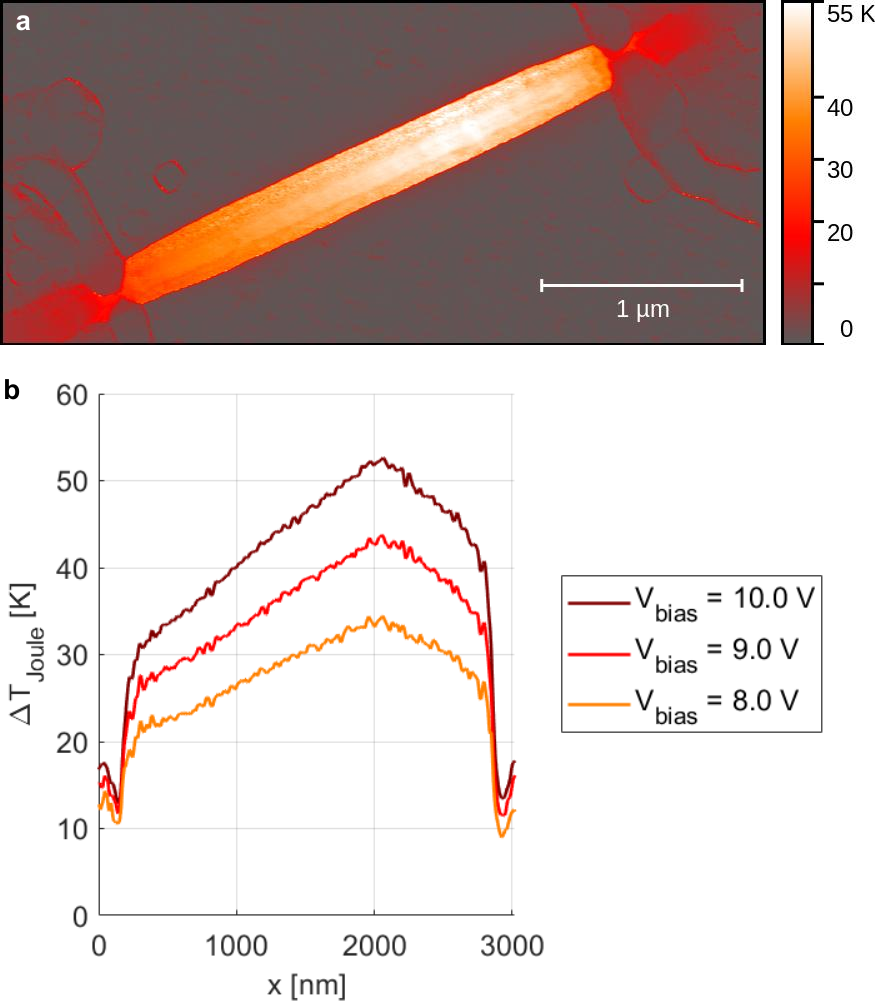}
	\caption[A floating figure]{Joule heating in an operando c-Al wire. a) shows the heat map measured with the SThM and b) shows the temperature profile along the nanowire axis under different voltage bias.}
	\label{fig:AlJoule} 
\end{figure}

Thermal measurements are also conducted on a c-Al NW. However, the electrical characterization indicates leaking of the device. Nonetheless,
the thermal measurements give some insight on where heat is generated. The Joule heat map in figure \ref{fig:AlJoule} shows heating of the entire wire, compared to the Al-Ge-wire, where it is more localized around the Ge-segment. Even more striking is the heating of the contacts which is not present in the heterostructure. In the profile lines along the wire, also shown in figure A.4, a distinct heat source is identified. Most probably it is located where the Ge-segment disappeared during fabrication, leaving behind some grain boundary or impurities inducing more scattering of electrons during operation. Consequently, the electrical conductivity of the single crystalline Al-parts of the heterostructure wire is expected to be higher than the values determined for the entire c-Al-wires due to the absence of this defect.

\bibliography{bibliography.bib}